\documentclass{article}
\usepackage{spconf,amsmath,amssymb,graphicx}
\usepackage[ruled,vlined]{algorithm2e}
\usepackage{subfigure}
 \usepackage[below]{placeins}
\usepackage{relsize,balance}

 \usepackage{cite}

\usepackage{color}  

\newcommand{\cb}[1]{{\boldsymbol{#1}}}
\newcommand{\cp}[1]{\ifmmode {\mathcal{#1}}\else ${\mathcal{#1}}$\fi}

\newcommand{\bpsi}{\boldsymbol{\psi}}

\newcommand{\br}{\boldsymbol{r}}

\newcommand{\bw}{\boldsymbol{w}}
\newcommand{\bx}{\boldsymbol{x}}

\newcommand{\bv}{\boldsymbol{v}}

\newcommand{\bB}{\boldsymbol{B}}
\newcommand{\bC}{\boldsymbol{C}}
\newcommand{\bK}{\boldsymbol{K}}
\newcommand{\bQ}{\boldsymbol{Q}}
\newcommand{\bG}{\boldsymbol{G}}
\newcommand{\bP}{\boldsymbol{P}}
\newcommand{\bH}{\boldsymbol{H}}
\newcommand{\bA}{\boldsymbol{A}}
\newcommand{\bR}{\boldsymbol{R}}

\newcommand{\bsig}{\boldsymbol{\sigma}}

\newcommand{\bGam}{\boldsymbol{\Gamma}}
\newcommand{\bI}{\boldsymbol{I}}

\newcommand{\N}[1]{\cp{N}_{#1}}
\newcommand{\C}{\cp{C}}

\newcommand{\vc}{\text{vec}}

\newcommand*{\Scale}[2][4]{\scalebox{#1}{$#2$}}%
\newcommand{\ssum}[2]{\Scale[1.01]{\sum\limits_{\mathsmaller{#1}}^{\mathsmaller{#2}}}}

\newtheorem{theorem}{Theorem}

\title{Diffusion LMS for Clustered Multitask Networks}
\name{Jie Chen $^{\star}$ \qquad C\'edric Richard $^{\star}$ \qquad Ali H. Sayed $^{\dagger}$ \thanks{This work was partly supported by the CNRS, France (Display project, Mastodons). {The} work of A. H. Sayed was supported in part by NSF grant CCF-1011918.}}
\address{$^{\star}$ Universit\'e de Nice Sophia-Antipolis, CNRS, France\\
                  $^{\dagger}$ University of California, Los Angeles, USA \\
                  \{jie.chen, cedric.richard\}@unice.fr \qquad sayed@ee.ucla.edu  }

\begin{document}
\ninept
\maketitle
\begin{abstract}
Recent research works on distributed adaptive networks have intensively studied the case where the nodes estimate a {common} parameter vector collaboratively. However, there are many applications that are multitask-oriented in the sense that there are multiple parameter vectors {that need to be} inferred simultaneously. In this paper, we employ diffusion strategies to develop distributed algorithms that address clustered multitask problems by minimizing an appropriate mean-square error criterion with $\ell_2$-regularization. Some results {on the mean-square} stability and convergence of the algorithm are also provided. Simulations are conducted to {illustrate} the theoretical findings.
\end{abstract}
\section{Introduction}
\vspace{-2mm}
Distributed adaptive learning {is an} attractive and challenging {subject} within the area of multi-agent networks.  {It leads to algorithms that are able to continuously adapt and learn, and that are particularly suitable for tracking concept drifts in the measured data.  The resulting distributed algorithms offer an important alternative to centralized  solutions with advantages resulting from scalability, robustness, and decentralization}. Several useful distributed strategies for online parameter estimation {have been proposed in the literature}, including consensus strategies~\cite{Nedic2009,Kar2009,Srivastava2011}, incremental strategies~\cite{Bertsekas1997,Rabbat2005,Blatt2007,Lopes2007incr}, and diffusion strategies~\cite{Sayed2013diff,Sayed2013intr,Lopes2008diff,Cattivelli2010diff,ChenUCLA2012,ChenUCLA2013}. {Incremental techniques require the determination of a cyclic path that runs across all nodes, which is generally an NP-hard problem. Besides, incremental solutions are sensitive to link failures. On the other hand, diffusion strategies are attractive since they are scalable, robust, and enable continuous adaptation and learning. In addition, for data processing over adaptive networks, diffusion strategies have been shown to have superior stability and performance ranges~\cite{Tu2012} than consensus-based implementations. Accessible overviews of recent results on diffusion adaptation can be found in~\cite{Sayed2013diff,Sayed2013intr}. }


An inspection of the literature on distributed algorithms shows that most {existing} works focus primarily, though not exclusively~\cite{Zhao2012,Tu2012decision,Bogdanovic2013}, on the case where the nodes have to estimate a single parameter vector collaboratively. {We refer to problems of this type as  \emph{single-task} problems}. However, many problems of interest happen to be \emph{multitask}-oriented in the sense that there are multiple parameter vectors to be inferred simultaneously and in a collaborative manner. Multitask learning problems have been studied by the machine learning community in several contexts, {including web page categorization~\cite{chen2009mtl}, web-search ranking~\cite{Chapelle2011mtl}, disease progression modeling~\cite{zhou2011mtl}, among other areas}. Clearly, this concept {is also relevant in the context of estimation over adaptive networks. Initial investigations along these lines for the traditional diffusion strategy appear in~\cite{Zhao2012,chen2013performance}}. In this article, we consider the situation where there are connected clusters of nodes, and each cluster has a parameter vector to estimate. The estimation still needs to be {performed cooperatively} across the network because the data across the clusters may be correlated and, therefore, cooperation across clusters can be beneficial. The aim of this paper is to derive a diffusion strategy that is able to solve the clustered multitask estimation problem, and to provide  analytical results for convergence in terms of mean weight error and mean-square error.

{\textbf{Notation}}.  Small letters $x$ denote scalars, and boldface small letters $\bx$ denote column vectors. Boldface capital letters $\bR$ represent matrices, and {the} operator $(\cdot)^\top$ {denotes} matrix transposition. $\bI_N$ denotes {the} $N\times N$ identity matrix. $\N{k}$ denotes the neighbors of node $k$, including $k$, whereas $\N{k}^-$ denotes the neighbors of node $k$, excluding $k$. $\C_i$ is the cluster $i$, i.e., index set of nodes in the $i$-th cluster. $\C(k)$ denotes the cluster to which node $k$ belongs. Finally, $\otimes$ denotes the Kronecker product, and $\text{vec}(\cdot)$ stacks the columns of a matrix {on top of each other} into a vector.

\section{Network model and problem formulation}
\vspace{-2mm}
\subsection{Clustered multitask network}
\vspace{-2mm}
Consider a connected network consisting of $N$ nodes. The problem is to estimate an $L\times 1$ unknown vector at each node $k$ from collected data. Node $k$ has access to time sequences $\{d_k(n), \bx_k(n)\}$, with $d_k(n)$ {representing} the reference signal, and $\bx_k(n)$ {denoting} an $L\times 1$ regression vector  with {covariance} matrix $\bR_{x,k}=E\{\bx_k(n)\bx_k^\top(n)\} >0$. The data at node $k$ are assumed to be related via the linear model:
\begin{equation}
          \label{eq:datamodel}
           d_k(n) = \bx_k^\top(n)\,\bw_k^\star + z_k(n)
\end{equation}
where $\bw_k^\star$ is an unknown parameter vector at node $k$, and $z_k(n)$ is a zero-mean, i.i.d. noise that is independent of every other signal and has variance $\sigma^2_{z,k}$. We assume that there are {$Q$ clusters and, therefore, $Q$ tasks} to be performed. We also assume that the nodes in {the} same cluster perform the same estimation task. The optimum parameter vectors $\bw_k^\star$ are constrained to be equal within each cluster, but similarities between neighboring clusters are allowed to exist, namely,
\begin{align}
           \bw_k^\star = \bw_{\C_q}^\star \qquad &\text{for } \forall k \in \C_q \\
           \bw_{\C_p}^\star \sim \bw_{\C_q}^\star \qquad &\text{if } \C_p, \; \C_q \text{ are connected}
\end{align}
where $p$ and $q$ denote two cluster indexes, and $\sim$  represents a similarity relationship in some sense. The reader is referred to Fig.~\ref{fig:Simu1A} for an illustration {showing a network with $N=15$ nodes and $Q=3$ clusters.}

\subsection{Problem formulation}
\vspace{-2mm}
Clustered multitask networks require that nodes  in the same cluster estimate the same coefficient vector. 
{We associate a} mean-square error cost function, $J_k(\bw_{\C(k)})$,  with each node $k$ such that
\begin{equation}
            \label{eq:MSEk}
            J_k(\bw_{\C(k)}) = E\big\{|d_k(n) - \bx_k^\top(n)\bw_{\C(k)}|^2\big\}.
\end{equation}
In order to promote similarities among adjacent clusters, appropriate regularization can be used. In this paper, we simply introduce the squared $\ell_2$-norm as a possible regularizer, namely,
\begin{equation}
          \label{eq:regk}
         \Delta(\bw_{\C(k)}, \bw_{\C(\ell)}) = \|\bw_{\C(k)} - \bw_{\C(\ell)}\|^2.
\end{equation}
Combining~\eqref{eq:MSEk} and~\eqref{eq:regk} yields the following regularization problem at the level of the entire network:
\begin{equation}
         \label{eq:MSEl2}
        \begin{split}
                 \overline{J^{\text{glob}}}(\bw_{\C_1}, \dots, &\bw_{\C_Q}) =\ssum{k=1}{N}\, E\big\{ |d_k(n) - \bx_k^\top(n)\,\bw_{\C(k)} |^2\big\}  \\
                   &+ \frac{\tau}{2} \,  \ssum{k=1}{N} \; \ssum{\ell\in\N{k} \backslash \C(k)}{}   \rho_{k\ell} \,\|\bw_{\C(k)} - \bw_{\C(\ell)}\|^2
          \end{split}
\end{equation}
where the second term on the RHS of expression~\eqref{eq:MSEl2} promotes similarities between neighboring clusters, with non-negative strength parameter $\tau$ and non-negative weights $\rho_{k\ell}$. 
{We seek a distributed solution to~\eqref{eq:MSEl2}. For that purpose, we first associate with} the $i$-th cluster, the following cost function
\begin{equation}
         \label{eq:MSEl2cls}
        \begin{split}
                 \overline{J_{\C_i}}&(\bw_{\C_i}) = \ssum{k\in\C_i}{} E\big\{ |d_k(n) - \bx_k^\top(n)\,\bw_{\C(k)} |^2\big\}  \\
                   &+ \frac{\tau}{2} \,  \ssum{k\in\C_i}{} \ssum{\ell\in\N{k} \backslash \C(k)}{} (\rho_{k\ell}+\rho_{k\ell}) \,\|\bw_{\C(k)} - \bw_{\C(\ell)}\|^2
          \end{split}
\end{equation}
Note that for given $\bw_{\C(\ell)}$ with $\ell\in\N{k} \backslash \C(k)$, {the costs in}~\eqref{eq:MSEl2} and~\eqref{eq:MSEl2cls} have the same gradient {vectors} relative to $\bw_{\C_i}$.  In order that each node can solve the problem autonomously and adaptively using only local interactions, we {shall} derive a distributed iterative algorithm for solving~\eqref{eq:MSEl2} by considering~\eqref{eq:MSEl2cls} since both cost functions have the same gradient {information}.

\section{Distributed adaptive estimation algorithm}
\vspace{-2mm}
\subsection{Local cost decomposition and problem relaxation}
\vspace{-2mm}
{We first note that a steepest-descent solution that is based on~\eqref{eq:MSEl2cls} will require every node in the network to have access to the statistical second-order moments of the data over its cluster. There are two problems with this scenario. First, nodes can only have access to information from their immediate neighborhood and the cluster of every node $k$ may include nodes that are not direct neighbors of $k$. Second, nodes rarely have access to the data statistical moments; instead, they have access to data generated from distributions with these moments. Therefore, more is needed to enable a distributed solution that relies solely on local interactions within neighborhoods and that relies on measured data as opposed to statistical moments. To derive a distributed algorithm, we follow the approach of~\cite{Sayed2013intr,Cattivelli2010diff}. The first step in this approach is to show how to express the cost~\eqref{eq:MSEl2cls} in terms of other local costs that only depend on data from neighborhoods.} 

{We start by introducing an} $N\times N$ right stochastic matrix $\bC$ with non-negative entries $c_{\ell k}$ such that
\begin{equation}
\vspace{-2mm}
            \ssum{k=1}N c_{\ell k} = 1, \quad \text{and} \quad c_{\ell k} =0 \,\text{ if }\, k\notin \cp{N}_{\ell} \cap \C(\ell).
\end{equation}
With these coefficients, we associate a local cost function of the following form {with} each node $k$:
\begin{equation}
             \label{eq:Jloc}
            J_k^{\text{loc}}(\bw_{\C(k)}) = \ssum{\ell\in\N{k}\cap\C(k)}{} c_{\ell k} E \big\{|d_\ell(n) - \bx_\ell^\top(n)\bw_{\C(k)}|^2\big\}
\end{equation}
In~\eqref{eq:Jloc}, note that $\bw_{\C(k)} = \bw_{\C(\ell)}$ because $\ell\in\C(k)$. To make the notation simpler, we shall write $\bw_k$ instead of $\bw_{\C(k)}$, {and} consequently $\bw_k = \bw_\ell$ for all $\ell\in\C(k)$. To take interactions {among} neighboring clusters into account, we modify~\eqref{eq:Jloc} by associating a regularized local cost function {with} node $k$ of the following form
\begin{equation}
	\label{eq:Jlocr}
	\begin{split}
	\overline{J_k^{\text{loc}}}(\bw_k) &= \ssum{\ell\in\N{k}\cap\C(k)}{}  c_{\ell k} 	\,E \big\{|d_\ell(n) - \bx_\ell^\top(n)\,\bw_k|^2 \big\}  \\
	           &+ \frac{\tau}{2} \ssum{\ell\in\N{k}\backslash\C(k)}{} (\rho_{k\ell}+\rho_{\ell k}) \|\bw_k - \bw_\ell\|^2.
	\end{split}
\end{equation}
{Observe that this local cost is now solely defined in terms of information that is available to node $k$ from its neighbors. It can then be verified that the} following  relation between~\eqref{eq:Jlocr} and~\eqref{eq:MSEl2cls} holds:
\begin{equation}
	\label{eq:JclusterSum}
	\begin{split}
	\overline{J_{\C(k)}}(\bw_k) 
	 =  \overline{J_k^{\text{loc}}}(\bw_k) + \ssum{\ell\in\C(k) \backslash k }{} \overline{J_\ell^{\text{loc}}}(\bw_\ell)
	 \end{split}
\end{equation}
Let $\bw_k^o$ denote the minimizer of the local cost~\eqref{eq:Jlocr}, given $\bw_{\ell}$ for all $\ell\in\N{k}\backslash\C(k)$. A completion-of-squares argument shows that, for any $k$, the cost $\overline{J_k^{\text{loc}}}(\bw_k)$ can be expressed as
\begin{equation}
	\label{eq:local.reg.eqv}
	\overline{J_k^{\text{loc}}}(\bw_k) =  \overline{J_k^{\text{loc}}}(\bw_k^o)
	+ \|\bw_k - \bw_k^o\|^2_ {\overline{\bR}_{k}}
\end{equation}
{where}
 \begin{equation}  
         \vspace{-2mm} 
         \overline{\bR}_{k}=\ssum{\ell\in\N{k}\cap\C(k)}{} c_{\ell k}\,\bR_{x, \ell} + \frac{\tau}{2}\ssum{\ell\in\N{k}\backslash\C(k)}{} (\rho_{k\ell}+\rho_{\ell k})\bI_L.
 \end{equation}
Substituting~\eqref{eq:local.reg.eqv} into the second term on the RHS of~\eqref{eq:JclusterSum}, and discarding the terms depending on $\bw_k^o$ since they are independent of the optimization variables in the cluster, we can consider the following equivalent alternative to \eqref{eq:JclusterSum} at node $k$:
 \begin{equation}
	\label{eq:Jglob.eqv}
         \overline{J_{\C(k)}}(\bw_{k}) = \overline{J_k^{\text{loc}}}(\bw_k)+\ssum{\ell \in\C(k)\backslash k}{}  \|\bw_\ell - \bw^o_\ell\|^2_{\overline{\bR}_{\ell}}
 \end{equation}
 where it holds that $\bw_k = \bw_\ell$  because $\ell \in \C(k)$. Therefore, the gradient of~\eqref{eq:Jglob.eqv} with respect to $\bw_k$ is equivalent to that of~\eqref{eq:MSEl2cls}. However, the second term of~\eqref{eq:Jglob.eqv} still requires multi-hop information passing. In order to avoid this situation, we relax~\eqref{eq:Jglob.eqv} at node $k$ by considering only information originating form its neighbors, i.e.,
  \begin{equation}
	\label{eq:Jglob.eqv.relax}
		\overline{J_{\C(k)}}'(\bw_k) = \overline{J_k^{\text{loc}}}(\bw_k)+\ssum{\ell\in \N{k}^-\cap\C(k)}{}   \|\bw_k - \bw^o_\ell\|^2_{\overline{\bR}_{\ell}}.
 \end{equation}
Usually, the weighting matrices $\overline{\bR_{\ell}}$ are unavailable. 
Following an argument based on the Rayleigh-Ritz characterization of eigenvalues, a useful strategy is to replace each matrix $\overline{\bR_{\ell}}$ by a weighted multiple of the identity matrix, say, as
\begin{equation}
	\|\bw_k - \bw^o_\ell\|^2_{\overline{\bR}_{\ell}} \approx {b_{\ell k}} \, \|\bw_k - \bw^o_\ell\|^2
\end{equation}
The coefficients $b_{\ell k}$ will be incorporated into a left stochastic matrix to be defined and, therefore, the designer does not need to worry about the selection of {these coefficients} at this stage~\cite{Sayed2013intr}. Based on the arguments presented so far,  {expression}~\eqref{eq:Jglob.eqv.relax} can then be relaxed to the following form:
 \begin{equation}
	\label{eq:Jglob.eqv.relax2}
         \begin{split}
         &\overline{J_{\C(k)}}''(\bw_k) = \ssum{\ell\in\N{k}\cap\C(k)}{} c_{\ell k} \, E\big\{|d_\ell(n) - \bx_\ell^\top(n)\,\bw_k|^2\big\} \\
         	&+ \frac{\tau}{2}  \ssum{\ell\in\N{k}\backslash\C(k)}{}\hspace{-3mm} (\rho_{k\ell}+ \rho_{\ell k}) \,\|\bw_k - \bw_\ell\|^2  +\hspace{-3mm} \ssum{\ell\in \N{k}^-\cap\C(k)}{}\hspace{-4mm} {b_{\ell k}}\|\bw_k - \bw^o_\ell\|^2.  
         \end{split}
  \end{equation}
 We now use~\eqref{eq:Jglob.eqv.relax2} to derive distributed strategies.

\subsection{Stochastic approximation algorithm}
\vspace{-2mm}
Let $\bw_k(n)$ denote the estimate for $\bw_k$ at iteration $n$. Using a constant step-size $\mu$ for each node, the update relation with {an} instantaneous approximation {for the gradient vector}, takes the following form:
\begin{equation}
	\label{eq:local.update}
	\begin{split}
		\bw_k(n+1) = &\bw_k(n) -\mu \hspace{-3mm} \ssum{\ell\in\N{k}\cap\C(k)}{} \hspace{-3mm} c_{\ell k} (\bx_\ell(n)^\top\bw_k(n) -d_{\ell}(n) ) \\
		                    &	- \mu\, \tau\hspace{-3mm}\ssum{\ell\in\N{k} \backslash \C(k)}{} \hspace{-3mm} \frac{\rho_{k\ell}+ \rho_{\ell k}}{2} \, \left(\bw_{k}(n)-\bw_{\ell}(n)\right) \\
		& - \mu\hspace{-3mm}\ssum{\ell\in{\N{k}^-\cap\C(k)}}{} \hspace{-3mm} b_{\ell k}\,\left(\bw_k(n)- \bw^o_\ell\right)
	\end{split}
\end{equation}
Among other possible forms,  expression~\eqref{eq:local.update} can be evaluated in two successive update steps:
\begin{align}
	&\bpsi_k(n+1)	= \bw_k(n) - \mu\Big[ \ssum{\ell\in\N{k}\cap\C(k)}{} \hspace{-3mm} c_{\ell k}(\bx_\ell(n)^\top\bw_k(n) -d_{\ell}(n) )   \nonumber \\
		&\hspace{1.7cm}+{\tau}\hspace{-3mm}\ssum{\ell\in\N{k} \backslash \C(k)}{} \hspace{-3mm} { \frac{\rho_{k\ell}+ \rho_{\ell k}}{2}}\, (\bw_{k}(n)-\bw_{\ell}(n))\Big] \label{eq:step1}\\
	&\bw_k(n+1)	= \bpsi_k(n+1) + \mu\hspace{-3mm}\sum_{\ell\in{\N{k}^-\cap\C(k)}}
		\hspace{-3mm} b_{\ell k}\,\left(\bw^o_\ell-\bw_k(n)\right)\label{eq:step2}
\end{align}
Following the same line of reasoning from~\cite{Sayed2013intr} in the single-task case,  we use $\bpsi_\ell(n+1)$ as a local estimate for  $\bw^o_\ell$ in~\eqref{eq:step2}, and replace $\bw_k(n)$ by $\bpsi_k(n+1)$. 
Step \eqref{eq:step2} then becomes
\begin{equation}
         \label{eq:wpsi}
	\bw_k(n+1) = \Big(1-  \mu \hspace{-3mm}  \ssum{\ell\in{\N{k}^-\cap\C(k)}}{} \hspace{-3mm} b_{\ell k} \Big) \bpsi_k(n+1) 
	+ \mu \hspace{-3mm} \sum_{\ell\in{\N{k}^-\cap\C(k)}} \hspace{-3mm} b_{\ell k}\, \bpsi_\ell(n+1).
\end{equation}
The coefficients in the above relation can be redefined as:
\begin{equation}
	\begin{split}
		a_{kk} 	& \triangleq 1-  \mu \!\! \ssum{\ell\in{\N{k}^-\cap\C(k)}}{} \!\! b_{\ell k}      \\
		a_{\ell k} 	& \triangleq \mu\, b_{\ell k}, \quad \ell\in\N{k}^-\cap\C(k)   \\
		a_{\ell k } 	& \triangleq 0, \quad  \ell\notin\N{k}\cap\C(k)
	\end{split}
\end{equation}
Let  $\bA$ be a left-stochastic matrix with ($\ell,k$)-th entry $a_{\ell k}$. With this notation, we arrive at the following adapt-then-combine (ATC) diffusion strategy for solving problem~\eqref{eq:MSEl2}:
\begin{equation}
	\label{eq:ATC_MSEl2}
     	\hspace{-3mm}\left\{
	\begin{split}
	&\bpsi_k(n+1) \!=\! \bw_k(n) \!+\! \mu \hspace{-2mm} \ssum{\ell\in\N{k}\cap\C(k)}{} \hspace{-2mm} c_{\ell k}[d_\ell(n)\!-\!\bx_\ell^\top(n)\bw_k(n)] \bx_\ell(n) \\
		& \qquad \qquad \qquad+ \tau\hspace{-3mm}\ssum{\ell\in\N{k} \backslash \C(k)}{} \hspace{-3mm}\frac{\rho_{k\ell}+\rho_{\ell k}}{2} \,(\bw_{\ell}(n) -  \bw_{k}(n))\\
	&\bw_k(n+1) \!=\! \hspace{-1mm} \ssum{\ell\in\N{k}\cap\C(k)}{} \hspace{-1mm} a_{\ell k} \, \bpsi_k(n+1)
	\end{split} \right.
	\vspace{-2mm}
\end{equation}

%

\section{Network performance analysis}
\vspace{-2mm}
In this section we examine the convergence properties and network performance of the adaptive diffusion strategy~\eqref{eq:ATC_MSEl2}. Let us denote by $\bw(n)$ and $\bw^\star$ the block weight estimate vector and the block optimum weight vector, respectively, both of size $L\times 1$, i.e.,
\begin{align}
     \bw(n) &= (\bw^\top_1(n), \dots, \bw^\top_N(n))^\top \\
     \bw^\star &= (\bw_1^{\star \top}, \dots, \bw_N^{\star\top})^\top
\end{align}
with $\bw^\star_k=\bw^\star_{\C(k)}$. Define the weight error vector by
\begin{equation}
      \bv(n) = \bw(n) - \bw^\star
\end{equation}
{Introduce} the block diagonal matrix $\bH = \text{diag}\left\{\bR_1,\,\dots,\bR_N\right\}$ with
\begin{equation}
	\bR_{k}=\sum_{\ell\in\N{k}\cap\C(k)} c_{\ell k}\,\bR_{x, \ell},
\end{equation}
{and let} $\bP$  be the matrix with $(k,\ell)$-th entry $\rho_{k\ell}$. {Introduce also} the block matrix 
\begin{equation}
	\bQ = \frac{1}{2}\left[\text{diag}\{(\bP+\bP^\top)\cb{1}\}-(\bP+\bP^\top)\right]\otimes \bI_L.
\end{equation}
and 
\begin{align}
       \bB &=  (\bA\otimes \bI_L)^\top \left[\bI_{LN} - \mu(\bH + \tau\bQ)\right] \\
      \br &=  (\bA\otimes \bI_L)^\top\bQ\,\bw^\star \\
      \bG &= (\bA\!\otimes\!\bI_L)^\top\! \bC_I^\top \text{diag}\{\sigma_{z,1}^2 \bR_{x,1},\dots,\sigma_{z,N}^2 \bR_{x,N}\}\bC_I(\bA\!\otimes\!\bI_L)
\end{align}
with $\bC_I = \bC\otimes \bI$.  Assume that the step-size $\mu$ is sufficiently small such that higher-order powers of $\mu$ can be neglected {and let}
\begin{equation}
	\label{eq:Kapp}
	\bK = \bB^\top \otimes \bB^\top.
\end{equation}
With these matrices and vectors, we have the following results~(proofs are omitted due to space constraints).
{\theorem (Stability in the mean) Assume data model~\eqref{eq:datamodel} {and that the regression data $\bx_k(n)$ is temporally white and independent over space}. Then, for any initial condition, the diffusion multitask strategy~\eqref{eq:ATC_MSEl2} asymptotically converges in the mean if the step-size is chosen to satisfy
\begin{equation}
	\label{eq:stepsize1}
	\begin{split}
		0 < \mu < \frac{2}{\max_k\{\lambda_{\max}(\bR_k)\}+{2}\tau\max_k\{\bQ_{kk}\}}
	\end{split}
\end{equation}
where $\lambda_\text{max}(\cdot)$ denotes the maximum eigenvalue of the matrix {arguement}. In addition, we have
\begin{equation}
	\label{eq:bias}
	\lim_{n\rightarrow\infty}E\{\bv(n)\} = \mu\tau( \bB- \bI_{LN} )^{-1}\br.
\end{equation}
}
{\theorem (Mean-square stability) Assume conditions in Theorem~1 hold. Then, the diffusion multitask strategy~\eqref{eq:ATC_MSEl2} is mean-square stable if the matrix $\bK$ is stable, which is guaranteed by sufficiently small step-sizes that also satisfy~\eqref{eq:stepsize1}.
}
{\theorem (Transient MSD) Considering a sufficiently small step-size $\mu$ that ensures mean and mean-square stability, the network  MSD learning curve, defined by $\zeta(n)=\frac{1}{N}E\{\|\bv(n)\|\}^2$, evolves according to  the following recursions for~$n\geq 0$:
\begin{equation}
	\begin{split}\label{eq:TransMSD}
		\zeta(n+1)  	&= \zeta(n) + \frac{1}{N} \Big( \mu^2\, \textnormal{\vc}(\bG^\top)^\top\, \bK^{n} \textnormal{\vc}(\bI_{LN}) \\
			&\hspace{-0.7cm}- E\{\|\bv(0)\|^2_{(\bI_{(NL)^2}-\bK)\bK^n\textnormal{\vc}(\bI_{LN})}\} +{\mu^2\tau^2} \|\br\|^2_{\bK^n\textnormal{\vc}(\bI_{LN})}  \\
					&\hspace{-0.7cm} -2\mu\tau\,(\bGam(n) + \left[ (\bB\,E\{\bv(n)\})^\top \otimes \br^\top \right] \textnormal{\vc}(\bI_{LN})\Big)   \\
		\bGam(n+1) &= \bGam(n)\, \bK +   \left[ (\bB\,E\{\bv(n)\})^\top \otimes \br^\top\right] (\bK-\bI_{(LN)^2})
    \end{split}
\end{equation}
with initial condition $\zeta(0) = \frac{1}{N}\|\bv(0)\|^2$ and $\bGam(0)  = \cb{0}_{(LN)^2}$.
}

{\theorem (Steady-state MSD) If convergence is achieved,  then the steady-state MSD  for the diffusion network~\eqref{eq:ATC_MSEl2} is given by
\begin{equation}
	\label{eq:MSD}
    \begin{split}
    	\zeta^*\!\!=\!\! \left[ \mu^2 \textnormal{\vc}(\bG^\top)^\top \!\!\! -\! 2\mu \tau ((\bB E\{\bv(\infty)\})^\top \otimes \br^\top)\right]  \sigma^o \!+\! \mu^2\tau^2   \|\br\|^2_{\bsig^o} 
    \end{split}
\end{equation}
where $\sigma^o=\frac{1}{N}(\bI_{(LN)^2} - \bK)^{-1}\textnormal{\vc}(\bI_{LN})$ and $E\{\bv(\infty)\}$ is determined by expression~\eqref{eq:bias}.
}

\section{Simulations}
\vspace{-3mm}

\subsection{Model validation}
\vspace{-2mm}

\begin{figure*}[t]
	\subfigure[Network topology.]{ \label{fig:Simu1A}
      		\includegraphics[trim = 10mm 17mm 12mm 15mm, clip, scale=0.35]{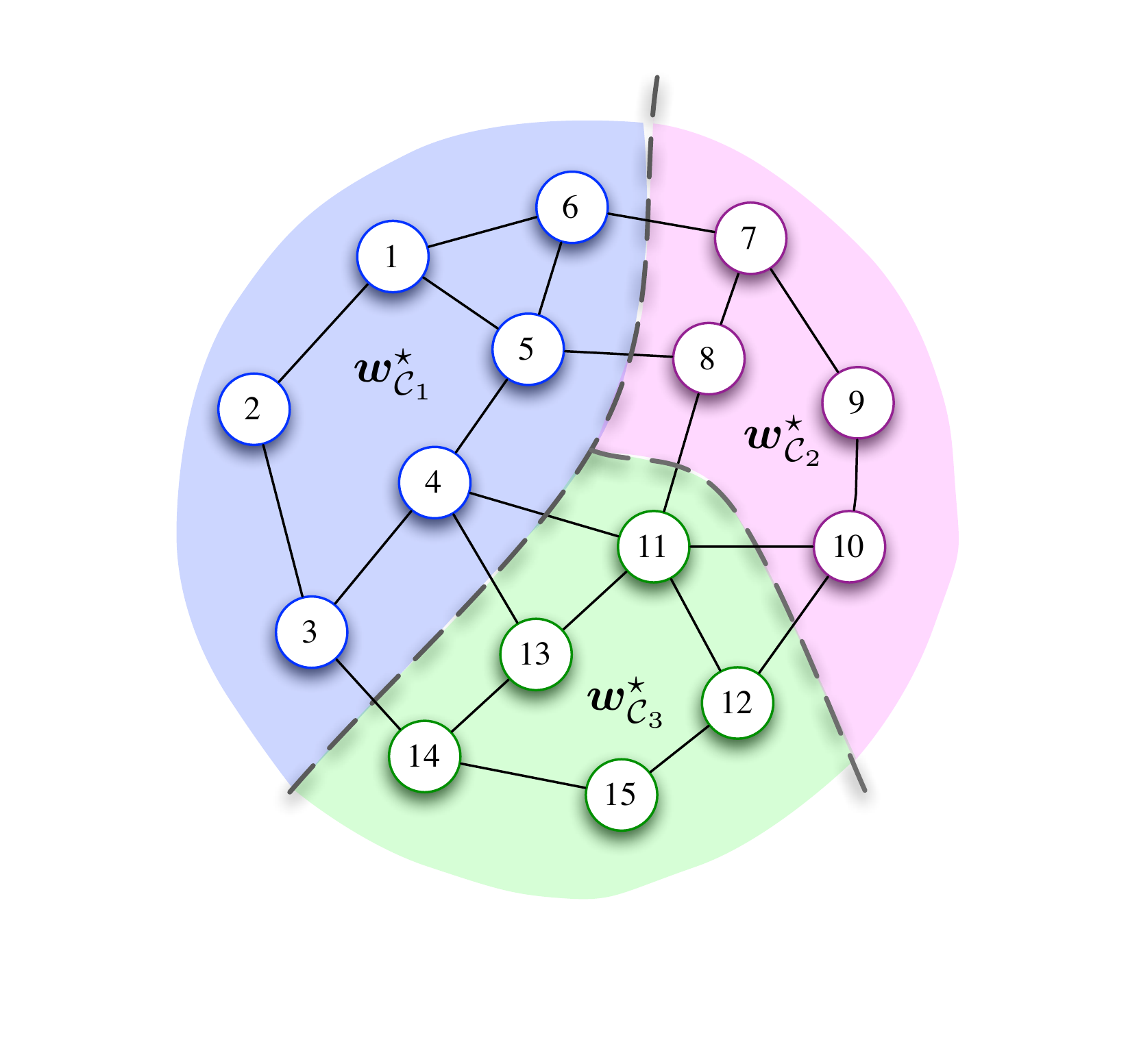} 
          }  
	\subfigure[Input and noise variances.]{  \label{fig:Simu1B}
      		\includegraphics[trim = 30mm 17mm 35mm 15mm, clip, scale=0.4]{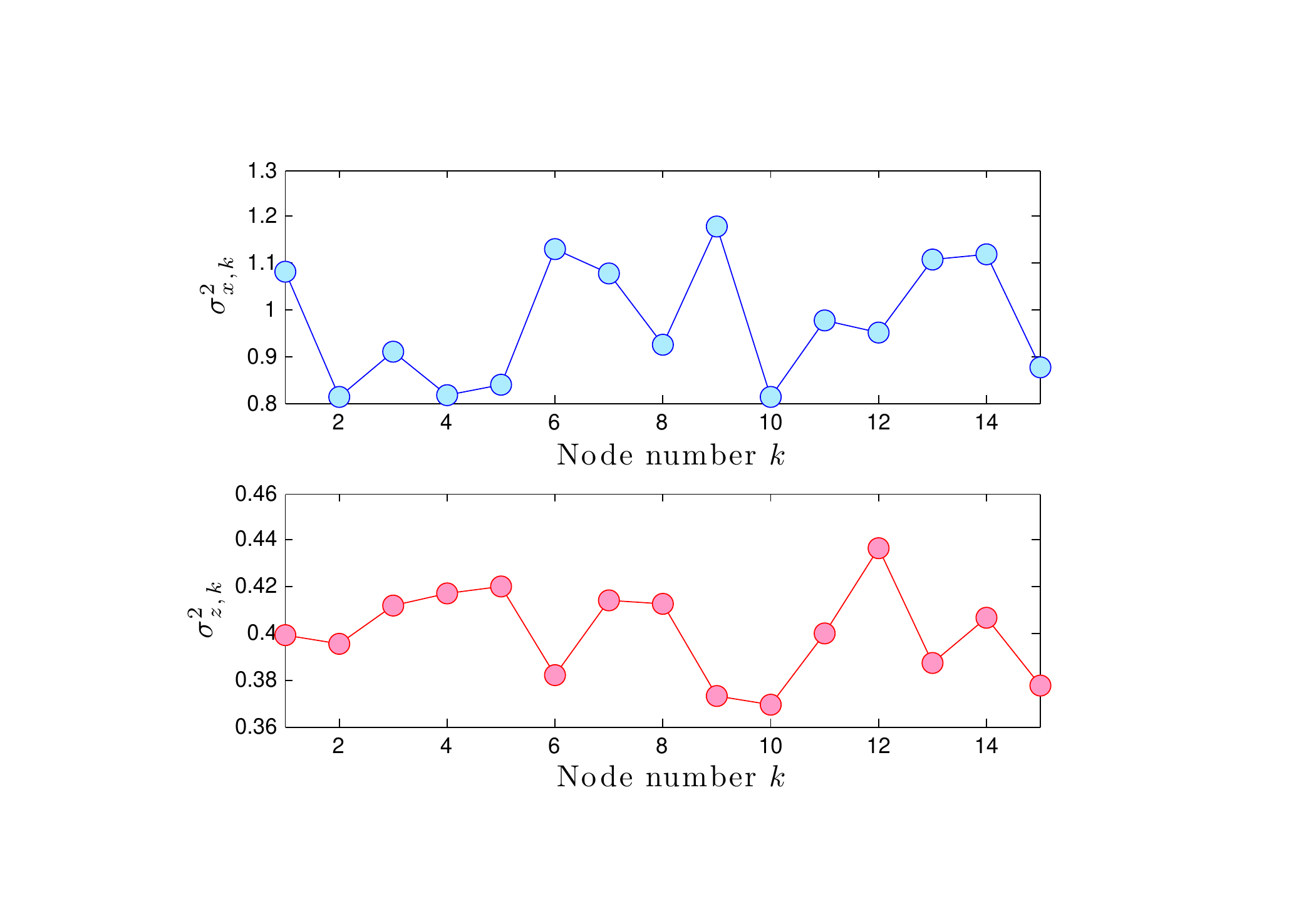}  
           }
 	\subfigure[Convergence illustration.]{  \label{fig:Simu1C}
      		\includegraphics[trim = 32mm 18.5mm 35mm 23mm, clip, scale=0.45]{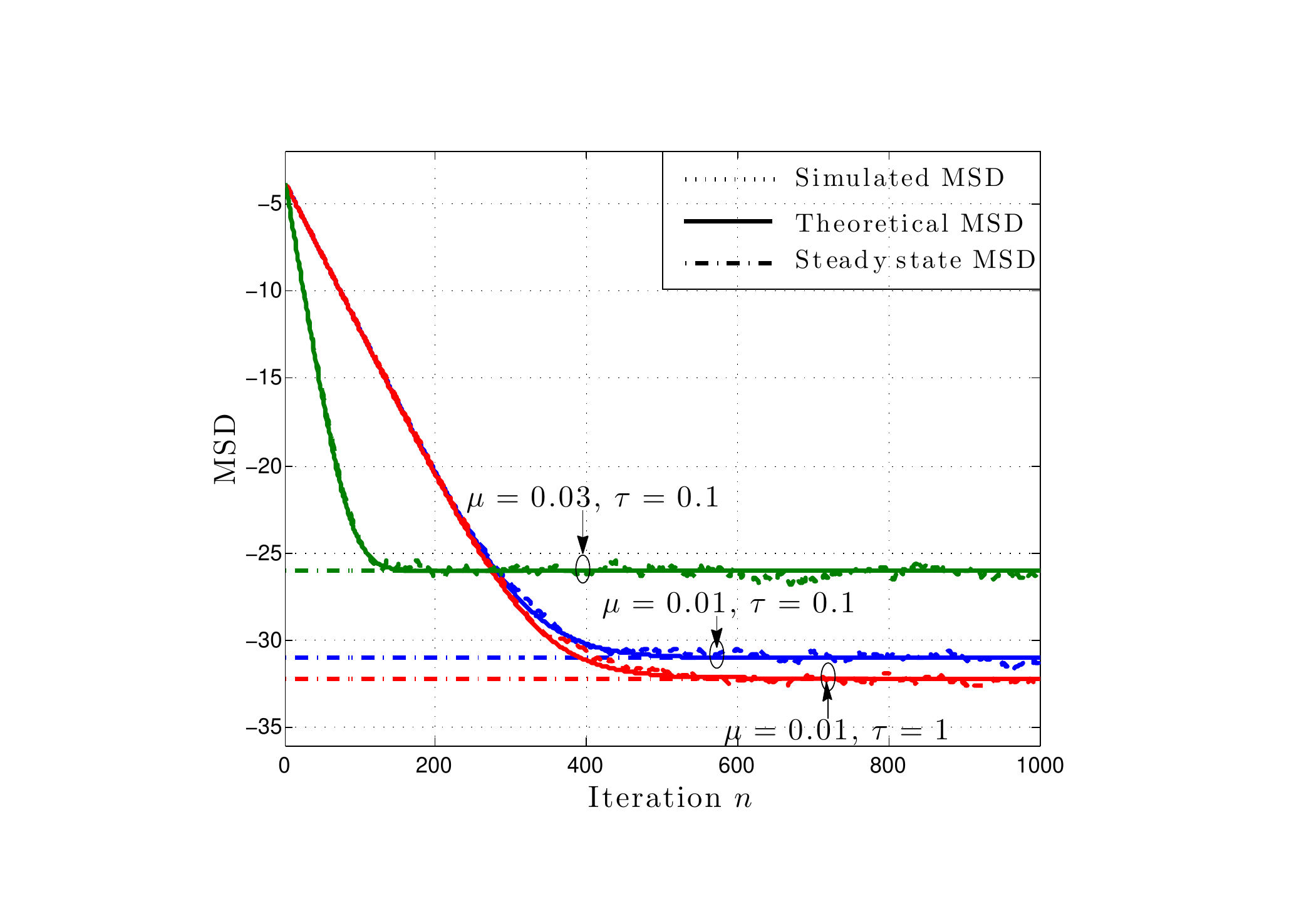}
        }
	\vspace{-0.5cm}
	\caption{Network configuration and result  illustration for Sec.~\ref{sec:expA}.}
	\label{fig:simu1}
	\vspace{-2mm}
\end{figure*}

\begin{figure*}
	\subfigure[Target locations.]{ \label{fig:Simu2A}
      		\includegraphics[trim = 38mm 14mm 38mm 23mm, clip, scale=0.4]{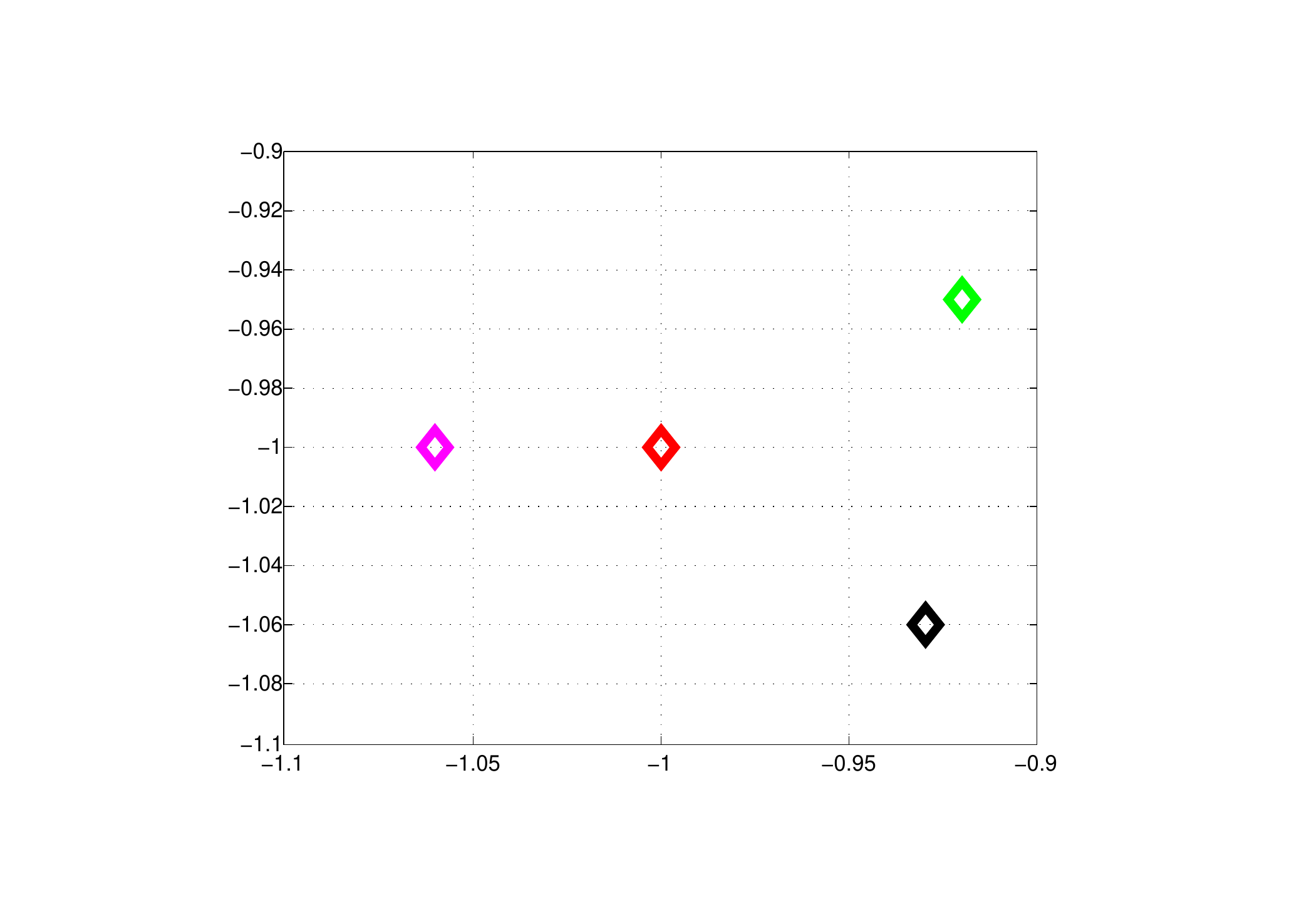}
          }  
	\subfigure[Network topology.]{  \label{fig:Simu2B}
      		\includegraphics[trim = 38mm 14mm 38mm 23mm, clip, scale=0.4]{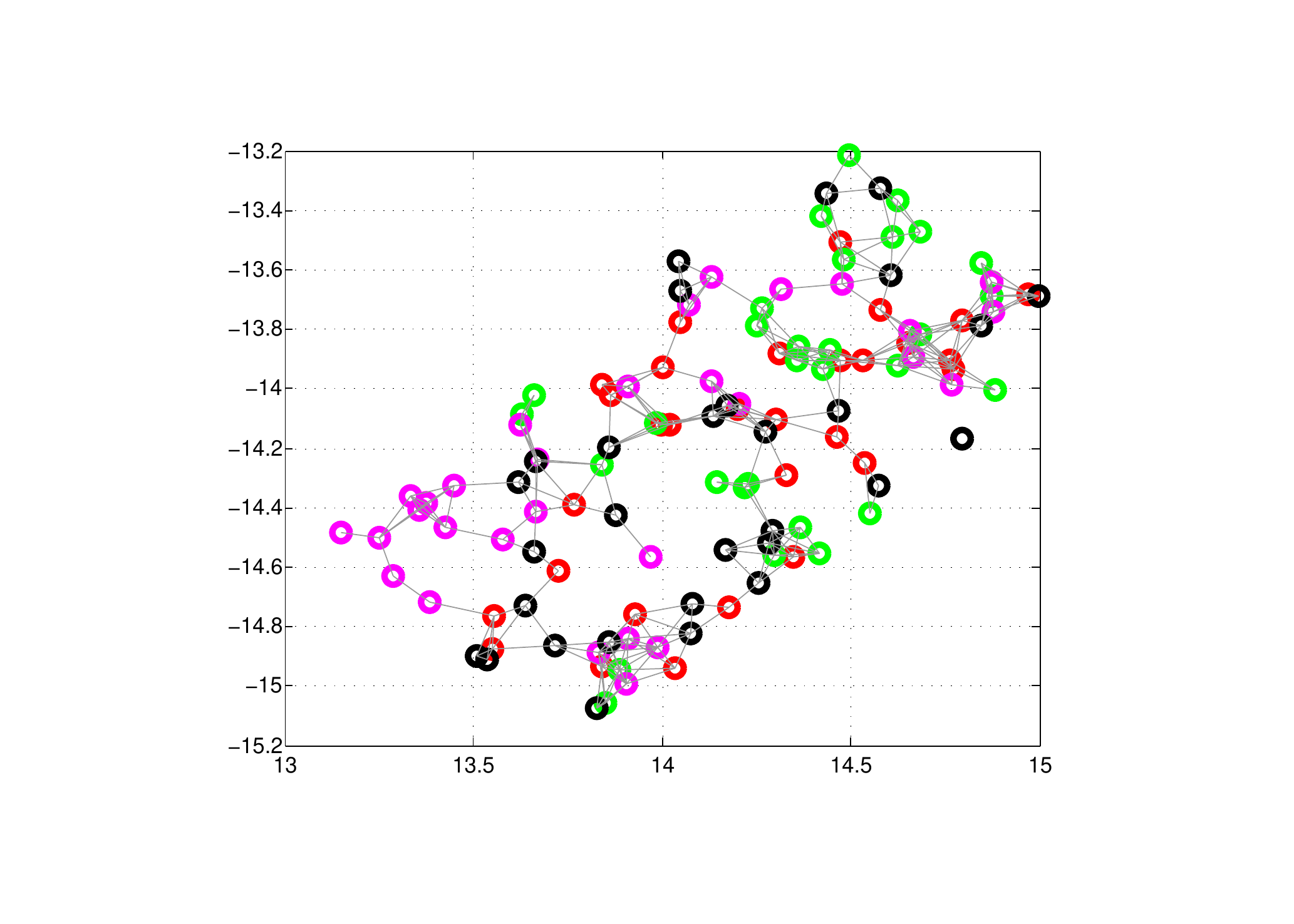} 
           }
 	\subfigure[MSD evolution.]{  \label{fig:Simu2C}
      		\includegraphics[trim = 31mm 20mm 36mm 23mm, clip, scale=0.45]{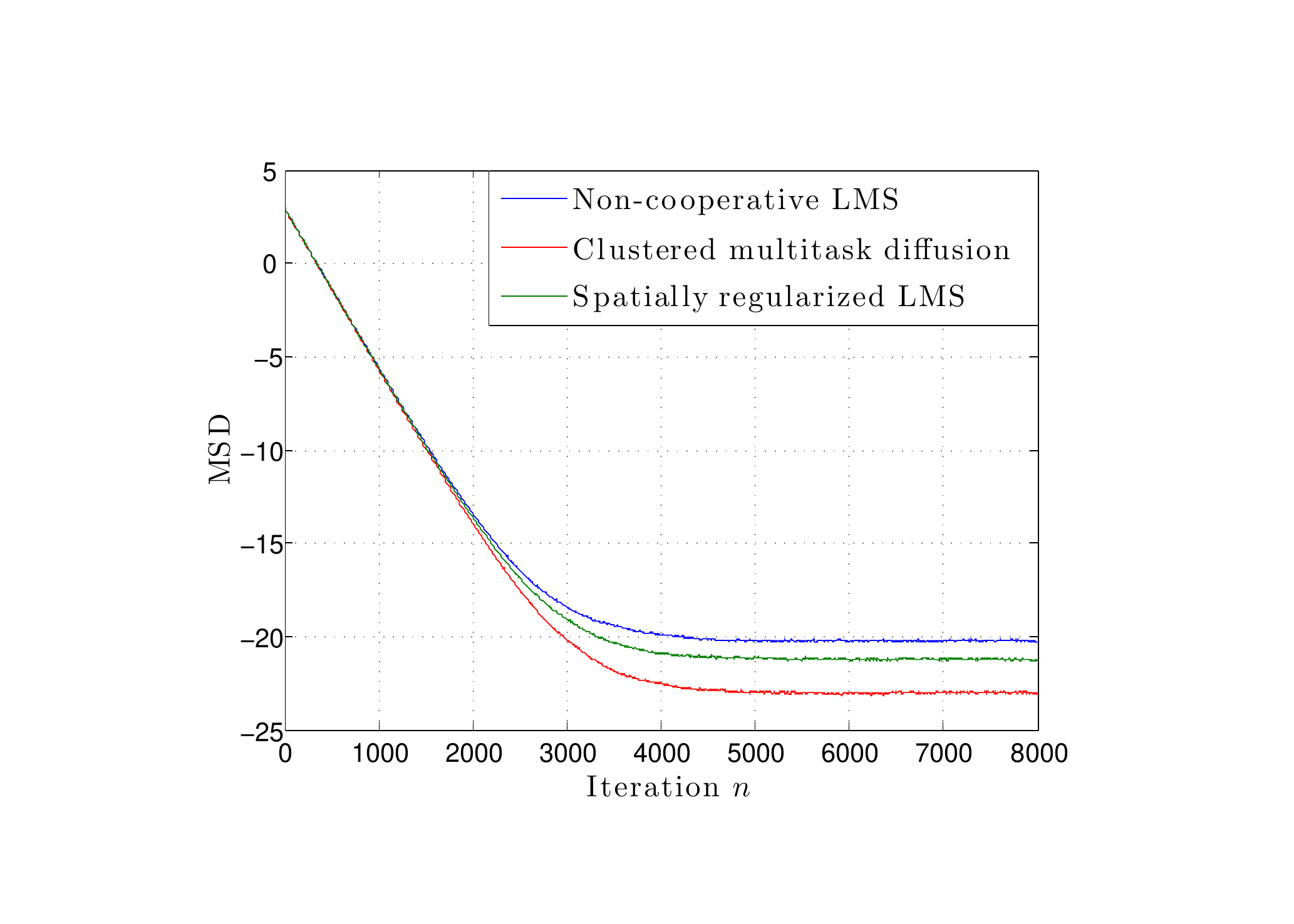}
        }
	\vspace{-5mm}
	\caption{Network configuration and result  illustration for Sec.~\ref{sec:expB}.}
\vspace{-0.5cm}
	\label{fig:simu2}
\end{figure*}

\label{sec:expA}
In this subsection we provide an illustrative example to show how the algorithm converges, and to {illustrate} theoretical models. We consider a network consisting of $15$ nodes with  connection and cluster structures shown in Fig.~\ref{fig:Simu1A}.  The parameter vectors to be estimated in each cluster are $\bw_{\C_1}^\star \!\!=\! (0.5238,-0.4008)^\top$, $\bw_{\C_2}^\star \!= \!(0.5065,-0.3965)^\top$ and $\bw_{\C_3}^\star \!=\! (0.4963,-0.3855)^\top$ respectively. Inputs $\bx(n)$ were zero-mean $2\times 1$ random vectors governed by a Gaussian distribution with covariance matrix $\bR_{x,k} = \sigma_{x,k}^2\bI_L$.  The noises $z_k(n)$ were i.i.d. zero-mean Gaussian random variables, independent of any other signal with variances $\sigma_{z,k}^2$. Variances $\sigma_{x,k}^2$ and $\sigma_{z,k}^2$ used in this experiment are depicted in Fig.~\ref{fig:Simu1B}.  Denoting the set cardinality by $|\cdot|$,  regularization weights $\rho_{k\ell}$ were uniformly chosen as $\rho_{k\ell} = |\N{k}\backslash\C(k)|^{-1}$ for $\ell\in\N{k}\backslash \C(k)$. We considered the diffusion algorithm with measurement diffusion governed by an identity matrix $\bC=\bI_N$, and a uniform combination matrix $\bA$ such that $a_{\ell k} = |\N{k}\cap\C(k)|^{-1}$ for $\ell \in \N{k}\cap\C(k)$.  The algorithm was run with different step sizes and regularization parameters $(\mu,\tau)$ such as $(0.01, 0.1)$, $(0.03, 0.1)$ and $(0.01, 1)$. Simulation results were obtained by averaging $100$ Monte-Carlo runs. Transient MSD curves were obtained by~\eqref{eq:TransMSD}. Steady-state MSD values were obtained by  expression~\eqref{eq:MSD}. Fig.~\ref{fig:Simu1C}  shows the evolutions of MSD and confirms theoretical analysis.

\vspace{-4mm}
\subsection{Multi-target localization}
\vspace{-2mm}
\label{sec:expB}
In this subsection we address an application of the  problem of multi-target localization. Existing localization methods based on the diffusion strategy assume point targets~\cite{Sayed2013intr}. However, in some situations, several distinct targets should be located.  In this simulation, the objective is to estimate coordinates of three nearby targets as shown in Fig.~\ref{fig:Simu2A} by a network composed by $120$ nodes, with approximately 20 distance units away from targets. Each node randomly selected a target $i\in\{1, 2, 3, {4}\}$. Nodes that selected the same target belong to the same cluster. The network connectivity and cluster structures are illustrated in  Fig.~\ref{fig:Simu2B}. Noise standard deviations were set to $\sigma_{\alpha,k} = 0.1$, $\sigma_{\beta,k}=0.01$ and $\sigma_{v,k}=0.3$ ({refer} to~\cite{Sayed2013intr} {for the interpretation of} these parameters). The proposed algorithm was run on each node with $\bC=\bI_N$, $a_{\ell k} = |\N{k}\cap\C(k)|^{-1}$ for $\ell \in \N{k}\cap\C(k)$, and  $\rho_{k\ell} = |\N{k}\backslash\C(k)|^{-1}$ for $\ell\in\N{k}\backslash \C(k)$. The step size was set {to $\mu = 0.1$}. The regularization strength was set to $\tau=0.01$. If each node is considered as a cluster, {then} algorithm~\eqref{eq:ATC_MSEl2} becomes a spatially regularized LMS, which was tested with the same parameter setting as the proposed algorithm. Non-cooperative LMS was also tested. MSD evolution curves were obtained by averaging over 100 Monte Carlo runs, as shown in Fig.~\ref{fig:Simu2C}. The benefit of cooperating and {clustering} is evidently illustrated.

\vspace{-3mm}

\section{Conclusion and perspectives}
\vspace{-2mm}
In this paper we derived a diffusion adaptation strategy for regularized learning over clustered multitask networks, and provided some convergence  properties of the algorithm. However it can be seen that due to the summation over all nodes by~\eqref{eq:MSEl2}, the problem inevitably leads to a symmetric regularization between pairs of nodes despite the fact that $\rho_{k\ell}\neq\rho_{\ell k}$. In order  to benefit from additional flexibility, we will study the asymmetric regularized learning multitask problem in  future work.

\vfill\pagebreak

\balance
\bibliographystyle{IEEEbib}
\bibliography{ref}

\begin{thebibliography}{10}

\bibitem{Nedic2009}
A.~Nedic and A.~Ozdaglar,
\newblock ``Distributed subgradient methods for multi-agent optimization,''
\newblock {\em IEEE Trans. Autom. Control}, vol. 54, no. 1, pp. 48--61, Jan.
  2009.

\bibitem{Kar2009}
S.~Kar and J.~M.~F. Moura,
\newblock ``Distributed consensus algorithms in sensor networks: Link failures
  and channel noise,''
\newblock {\em IEEE Trans. Signal Process.}, vol. 57, no. 1, pp. 355--369, Jan.
  2009.

\bibitem{Srivastava2011}
K.~Srivastava and A.~Nedic,
\newblock ``Distributed asynchronous constrained stochastic optimization,''
\newblock {\em IEEE J. Sel. Topics Signal Process.}, vol. 5, no. 4, pp.
  772--790, Aug. 2011.

\bibitem{Bertsekas1997}
D.~P. Bertsekas,
\newblock ``A new class of incremental gradient methods for least squares
  problems,''
\newblock {\em SIAM J. Optimiz.}, vol. 7, no. 4, pp. 913--926, Nov. 1997.

\bibitem{Rabbat2005}
M.~G. Rabbat and R.~D. Nowak,
\newblock ``Quantized incremental algorithms for distributed optimization,''
\newblock {\em IEEE J. of Sel. Topics Areas Commun.}, vol. 23, no. 4, pp.
  798--808, Apr. 2005.

\bibitem{Blatt2007}
D.~Blatt, A.~O. Hero, and H.~Gauchman,
\newblock ``A convergent incremental gradient method with constant step size,''
\newblock {\em SIAM J. Optimiz.}, vol. 18, no. 1, pp. 29--51, Feb. 2007.

\bibitem{Lopes2007incr}
C.~G. Lopes and A.~H. Sayed,
\newblock ``Incremental adaptive strategies over distributed networks,''
\newblock {\em IEEE Trans. Signal Process.}, vol. 55, no. 8, pp. 4064--4077,
  Aug. 2007.

\bibitem{Sayed2013diff}
A.~H. Sayed, S.-Y Tu, J.~Chen, X.~Zhao, and Z.~Towfic,
\newblock ``Diffusion strategies for adaptation and learning over networks,''
\newblock {\em IEEE Sig. Process. Mag.}, vol. 30, no. 3, pp. 155--171, May
  2013.

\bibitem{Sayed2013intr}
A.~H. Sayed,
\newblock ``Diffusion adaptation over networks,''
\newblock in {\em Academic Press Libraray in Signal Processing}, R.~Chellapa
  and S.~Theodoridis, Eds., pp. 322--454. Elsevier, 2013. Also available as
  arXiv:1205.4220 [cs.MA], May 2012.

\bibitem{Lopes2008diff}
C.~G. Lopes and A.~H. Sayed,
\newblock ``Diffusion least-mean squares over adaptive networks: Formulation
  and performance analysis,''
\newblock {\em IEEE Trans. Signal Process.}, vol. 56, no. 7, pp. 3122--3136,
  Jul. 2008.

\bibitem{Cattivelli2010diff}
F.~S. Cattivelli and A.~H. Sayed,
\newblock ``Diffusion {LMS} strategies for distributed estimation,''
\newblock {\em IEEE Trans. Signal Process.}, vol. 58, no. 3, pp. 1035--1048,
  Mar. 2010.

\bibitem{ChenUCLA2012}
J.~Chen and A.~H. Sayed,
\newblock ``Diffusion adaptation strategies for distributed optimization and
  learning over networks,''
\newblock {\em IEEE Trans. Signal Process.}, vol. 60, no. 8, pp. 4289--4305,
  Aug. 2012.

\bibitem{ChenUCLA2013}
J.~Chen and A.~H. Sayed,
\newblock ``Distributed {P}areto optimization via diffusion strategies,''
\newblock {\em IEEE J. Sel. Topics Signal Process.}, vol. 7, no. 2, pp.
  205--220, Apr. 2013.

\bibitem{Tu2012}
S.-Y. Tu and A.~H. Sayed,
\newblock ``Diffusion strategies outperform consensus strategies for
  distributed estimation over adaptive networks,''
\newblock {\em IEEE Trans. Signal Process.}, vol. 60, no. 12, pp. 6217--6234,
  Dec. 2012.

\bibitem{Zhao2012}
X.~Zhao and A.~H. Sayed,
\newblock ``Clustering via diffusion adaptation over networks,''
\newblock in {\em Proc. CIP}, Parador de Baiona, Spain, May 2012, pp. 1--6.

\bibitem{Tu2012decision}
S.-Y. Tu and A.~H. Sayed,
\newblock ``Adaptive decision making over complex networks,''
\newblock in {\em Proc. ASILOMAR}, Pacific Grove, CA. USA, Nov. 2012, pp.
  525--530.

\bibitem{Bogdanovic2013}
N.~Bogdanovi\'c, J.~Plata-Chaves, and K.~Berberidis,
\newblock ``Distributed incremental-based {LMS} for node-specific parameter
  estimation over adaptive networks,''
\newblock in {\em Proc. ICASSP}, Vancouver, Canada, May 2013, pp. 5425--5429.

\bibitem{chen2009mtl}
J.~Chen, L.~Tang, J.~Liu, and J.~Ye,
\newblock ``A convex formulation for leaning shared structures from muliple
  tasks,''
\newblock in {\em Proc. ICML}, Montreal, Canada, Jun. 2009, pp. 137--144.

\bibitem{Chapelle2011mtl}
O.~Chapelle, P.~Shivaswmy, K.~Q. Vadrevu, S.~Weinberger, Y.~Zhang, and
  B.~Tseng,
\newblock ``Multi-task learning for boosting with application to web search
  ranking,''
\newblock in {\em Proc. SIGKDD}, Washington DC, USA, Jul. 2010, pp. 1189--1198.

\bibitem{zhou2011mtl}
J.~Zhou, L.~Yuan, J.~Liu, and J.~Ye,
\newblock ``A multi-task learning formulation for predicting disease
  progression,''
\newblock in {\em Proc. SIGKDD}, San Diego, CA, USA, Aug. 2011, pp. 814--822.

\bibitem{chen2013performance}
J.~Chen and C.~Richard,
\newblock ``Performance analysis of diffusion {LMS} in multitask networks,''
\newblock in {\em Proc. IEEE CAMSAP}, Saint Martin, France, Dec. 2013, pp.
  1--4.

\end{thebibliography}

\end{document}